\documentclass[10pt,letterpaper]{article}
\usepackage[top=0.85in,left=1.5in,footskip=0.75in]{geometry}
\usepackage{amsmath,amssymb}

\usepackage{changepage}

\usepackage[utf8x]{inputenc}

\usepackage{textcomp,marvosym}

\usepackage{cite}

\usepackage{nameref,hyperref}

\usepackage{microtype}
\DisableLigatures[f]{encoding = *, family = * }

\usepackage[table]{xcolor}

\usepackage{array}

\newcolumntype{+}{!{\vrule width 2pt}}

\newlength\savedwidth

\usepackage[aboveskip=1pt,labelfont=bf,labelsep=period,justification=raggedright,singlelinecheck=off]{caption}

\makeatletter
\renewcommand{\@biblabel}[1]{\quad#1.}
\makeatother

\date{}

\usepackage{lastpage,fancyhdr,graphicx}
\usepackage{epstopdf}

\usepackage{rotating}
\usepackage{url}  
\usepackage{ifthen}  
\usepackage{multicol}   

\usepackage{float}

\usepackage{subcaption}
\usepackage[linesnumbered,ruled]{algorithm2e}
\usepackage{algorithmicx}
\usepackage{hyperref}
\usepackage{setspace,caption}
\usepackage[noend]{algpseudocode}
\algrenewcommand\algorithmicrequire{\textbf{Method:}}

\begin{document}
\vspace*{0.2in}

\begin{flushleft}
{\Large
\textbf\newline{\begin{myfont}MI-Sim\end{myfont}: A \begin{myfont}MATLAB\end{myfont} Package for the Numerical Analysis of Microbial Ecological Interactions}}
\newline
\\
Matthew J. Wade\textsuperscript{1*},
Jordan Oakley\textsuperscript{2},
Sophie Harbisher\textsuperscript{2},
Nicholas G. Parker\textsuperscript{2},
Jan Dolfing\textsuperscript{1}
\\
\bigskip
\textbf{1} School of Civil Engineering and Geosciences, Newcastle University, Newcastle-upon-Tyne NE1 7RU, United Kingdom
\\
\textbf{2} School of Mathematics and Statistics, Newcastle University, Newcastle-upon-Tyne NE1 7RU, United Kingdom
\\
\bigskip

*matthew.wade@ncl.ac.uk

\end{flushleft}

\section*{Abstract}

Food-webs and other classes of ecological network motifs, are a means of describing feeding relationships between consumers and producers in an ecosystem. They have application across scales where they differ only in the underlying characteristics of the organisms and substrates describing the system. Mathematical modelling, using mechanistic approaches to describe the dynamic behaviour and properties of the system through sets of ordinary differential equations, has been used extensively in ecology. Models allow simulation of the dynamics of the various motifs and their numerical analysis provides a greater understanding of the interplay between the system components and their intrinsic properties. We have developed the \begin{myfont}MI-Sim\end{myfont} software for use with \begin{myfont}MATLAB\end{myfont} to allow a rigorous and rapid numerical analysis of several common ecological motifs. \begin{myfont}MI-Sim\end{myfont} contains a series of the most commonly used motifs such as cooperation, competition and predation. It does not require detailed knowledge of mathematical analytical techniques and is offered as a single graphical user interface containing all input and output options. The tools available in the current version of \begin{myfont}MI-Sim\end{myfont} include model simulation, steady-state existence and stability analysis, and basin of attraction analysis. The software includes seven ecological interaction motifs and seven growth function models. Unlike other system analysis tools,  \begin{myfont}MI-Sim\end{myfont} is designed as a simple and user-friendly tool specific to ecological population type models, allowing for rapid assessment of their dynamical and behavioural properties.

\section*{Introduction}
Network motifs provide an approach to understand and characterise the behaviour of living systems at genomic, metabolic and ecological scales~\cite{milo02,bascompte05,levy15}. Food-webs, defined as a subset or module of larger, more complex networks, are used to analyse ecological interactions at the community or population level, as first described by mathematicians such as Lotka and Volterra, and have been widely used to explore phenomena observed at both macro- and micro-scales~\cite{bungay68,butler83,lauwerier86}. \\

Mathematical modelling of ecological interactions is affected by the model objective (e.g., observation, prediction, control), the availability of existing knowledge and data, and the structural complexity necessary to adequately describe the motif. For clarity, we define \textit{motif} here to be analogous to interaction \textit{modules} described by population ecologists, and the specific forms of these motifs are described widely in the literature (e.g.~\cite{Grosskopf14}). \\

The software presented here focuses on a mechanistic understanding of microbial interactions and, specifically, their analysis and simulation for two or three microbial species and associated substrates and products. The motif models are developed as systems of Ordinary Differential Equations (ODEs) used to describe the dynamics of and interactions between the individual organisms and their various components.  \\

Mathematical analysis of such model structures is commonplace in fields such as chemostat theory~\cite{aris77,kreikenbohm86,sivaprakash11,weedermann13}, predator-prey system analysis~\cite{fussmann00,gilpin72}, theoretical microbial ecology~\cite{xu11,wade16}, and more recently in application to synthetic microbiology~\cite{escalante15,zomorrodi16}. Methods that include steady-state analysis and basin of attraction characterisation are necessary to understand the stability, resilience and persistence of the modelled microbial populations. However, performing these analyses robustly requires a relatively high degree of competency with mathematical theory of dynamical systems. There are several tools available for the numerical analysis of dynamical ODEs. These include \begin{myfont}XPPAUT\end{myfont}~\cite{ermentrout02}, which is compiled in C, and \begin{myfont}MATCONT\end{myfont}~\cite{dhooge03}, a continuation software available for \begin{myfont}MATLAB\end{myfont}. However, whilst versatile, these tools are difficult for non-specialists to use without a background in modelling or mathematics. Furthermore, for use in systems with more than four ODEs, bifurcation and stability analysis is often problematic as finding exact solutions for higher-dimensional systems is non-trivial and often intractable. \\

We present here an analysis software, \begin{myfont}MI-Sim\end{myfont}, specifically for microbial communities. It provides a user-friendly environment in which microbiologists, microbial ecologists and biological mathematicians can rapidly and robustly characterise the dynamics of ecological motifs of up to three microbial species, without the requirement to develop their own code, construct models or have a detailed knowledge of the mathematics of dynamical systems analysis.

\section*{Mathematical analysis of ecological motifs}
\subsection*{Description of motifs}
Foremost, we aimed to develop a tool that enables users to model and analyse their own species interactions by making the software as generic as possible. Here, we have taken six common ecological motifs describing interactions between two distinct species, plus one extended motif that consists of three interacting species. The seven motifs, described in Table~\ref{motifs}, are simple networks commonly observed at both micro- and macro-scales, and provide a theoretical basis by which scientists can test hypotheses in suitably sized community networks~\cite{faust12,Song14,Grosskopf14,wade16}. 

\begin{table}[bp!]\caption{Description of ecological motifs available in the software. A negative interaction indicates a cost to the species, a positive interaction signifies the species gains a benefit and $0$ is a neutral interaction.}
\centering
\small
\begin{tabular}{llcc}
\hline 
Motif name & Ecological term & Interaction & No. species \\ \hline
Syntrophy  & Cooperation & $+/+$ & 2 \\
Food chain  & Commensalism & $0/+$  & 2 \\
Food chain with waste product inhibition  &  Predation & $-/+$  & 2    \\
No common metabolites  & No interaction & $0/0$  & 2  \\
Waste product inhibition  & Amensalism & $0/-$  & 2  \\
Substrate competition  & Competition & $-/-$  & 2  \\
Three species food-web  & Competition/Cooperation & $(+/-)/+/(-/+)$ & 3 \\ \hline
\end{tabular}
\label{motifs}
\end{table}

\subsection*{Development of the models}
\begin{myfont}MI-Sim\end{myfont} uses a deterministic rather than phenomenological approach for modelling and simulation of microbial species interactions. The described motifs are expressed as a series of ODEs, which describe the microbial growth, catabolic conversion processes, and species interactions within the system. The equations are developed using a standard mass-balance approach coupled with stoichiometric information describing the chemical transformation between reactants and products in the system. Whilst analytical approaches providing exact solutions are typically restricted to one or two species, numerical analysis allows extension to higher-dimensional models, albeit generating local rather than global solutions. \\

The models currently available in \begin{myfont}MI-Sim\end{myfont} take the following generalised form (shown here for one biomass and substrate pairing):\\
\begin{align}
\frac{\mathrm{d}S}{\mathrm{d}t} &= (S_{\mathrm{in}} - S)\alpha-\mu({S}){X} \\
\frac{\mathrm{d}X}{\mathrm{d}t} &= (\mu({S})-\alpha){X}
\end{align}
where $t$ is time, $X = X(t)$ is the biomass concentration, $S=S(t)$ is the substrate concentration, $S_{\mathrm{in}}$ is the influent substrate concentration, $\mu(S)$ is the substrate dependent growth rate of the biomass and $\alpha$ is the flow of substrate in the system. In chemostat theory, for example, $\alpha$ typically represents the dilution rate or nutrient exchange, which is the inflow of substrate divided by the bioreactor volume.\\

The seven motifs included in this version of the software have been developed according to this form. Users may manipulate the structure and behaviour of the different models by selecting appropriate parameter values. For example, all models contain a biomass decay rate term $k_{\mathrm{dec}}$, which may be removed by setting its value equal to zero. Bespoke models of $n$ coupled ODEs ($n$ being theoretically unlimited, although increasingly complex and larger systems may be restricted by CPU capacity), can be created by altering the code describing the default models, but this option is only recommended for users with some knowledge of \begin{myfont}MATLAB\end{myfont}. Future versions of the software will include a \textit{Model Builder} option to make this process more transparent.

\subsection*{Units}

Molarity ($\mathrm{M}$ or $\mathrm{mol/L}$) is the most commonly used unit for describing solution concentration in microbial systems. Kinetic studies of microorganism growth conventionally calculate parameters based on this unit for stoichiometric simplicity. However, in certain processes where solutes have an impact on the carbon oxidation state of the environment (e.g., wastewater treatment systems), solute strength is generally expressed in terms of Chemical Oxygen Demand (COD) for relevant components (i.e., those having a COD; $M$ is used for those with no COD~\cite{batstone02}). Accordingly, \begin{myfont}MI-SIM\end{myfont} allows for the selection of units to be used by the model.

\section*{Description of the software}

\subsection*{User interface}
The software is presented as a Graphical User Interface (GUI), thus removing any requirement for the user to be fluent with \begin{myfont}MATLAB\end{myfont} coding. The user-friendly environment can be easily navigated using toolbar options for selection of motif and mathematical analysis routine, as well as visualisation and reporting options. \\

A number of panels on the GUI enable the user to configure and parameterise the model and the simulation properties. The \textit{Growth Function} panel is used to select the growth model ($f_n$) in the model (default: Monod). Parameterisation of the model is done using the \textit{Parameter Values} panel. For dynamical analysis of the model the \textit{Initial Conditions} panel provides an interface to select the starting concentrations of the model variables (substrate, $S_n$ and biomass, $X_n$, where $n$ is the index of the substrate/species). The \textit{Solver Options} panel provides the user with a selection of in-built \begin{myfont}MATLAB\end{myfont} ODE solver methods (default:  \textit{ode23s} for stiff systems) and absolute and relative tolerance values. These tolerances act to control the error estimation of the algorithm for the given solver method (e.g., Rosenbrock formula of order two for \textit{ode23s}). It is not necessary for the user to understand the mechanics of the ODE solvers but, for non-trivial solutions, it may be necessary to adjust the tolerance values sacrificing speed for accuracy. A checkbox option \textit{Jacobian} may reduce the stiff solver run-time by supplying a sparse Jacobian matrix to the solver. This avoids costly calls to the rate of change function. The \textit{Simulation Options} panel is used to select parameters or variables and their value ranges for routines that analyse the models in a multiple-point simulation space (i.e., \textit{Multiple-point}, \textit{Basin of attraction}, and \textit{Phase portrait}). \\

The remaining panels are used to display and visualise the various outputs from the mathematical analysis of the motifs. The \textit{Equations} panel displays the set of differential equations for the selected motif. Running the \textit{multiple-point analysis} generates a heatmap as the algorithm calculates the Jacobian matrix for each parameter pair, and after completion a table showing the stability of each steady-state will be displayed. The \textit{Progress} panel displays information on percent completion, which is useful for monitoring routines requiring multiple iterations, together with some text on the task status. The \textit{Fixed Points \& Stability} panel displays the calculated fixed-point solutions and their stability for each variable. The output of this panel depends on the stability analysis selected in the \textit{Stability Check} panel and is only available for single-point analysis. Two plot windows are available to display graphically the numerical outputs. 

\subsection*{Analysis tools}
\begin{myfont}MI-Sim\end{myfont} comprises four numerical analysis tools used to fully characterise the motifs. They are presented here as pseudocode.

\subsubsection*{Single-point analysis}
This routine is used for analysing the dynamics and steady-state stability of the motif at a single point in the model parameter space, $\Theta$. The algorithm is summarised in Algorithm~\ref{algoa}. Briefly, the system of ODEs, $F(\tilde{x}(t))$, are solved algebraically to find the fixed-point solutions, $\tilde{x}^{g,*}$, where $g$ represents the number of fixed-points found. $\tilde{x}$ is the vector of variables ($X$ and $S$) associated with the motifs. Two methods are available for analysis of the stability of the fixed-points; linear stability analysis and the Routh-Hurwitz criterion. In the former, $\epsilon$ defines the tolerance threshold for stability. 

\begin{algorithm}
\KwIn{$\Theta, t, \tilde{x}(0), q, \epsilon$}
\KwOut{$\tilde{x}(t), \tilde{x}^{g,*} \in \mathbb{Re}^n \times \mathbb{Re}^g$ }
Substitute numerical parameters $\Theta$ into system of symbolic ODEs $F(\tilde{x}(t))$ \\
Solve $F(\tilde{x})$ numerically and plot solutions $\tilde{x}(t)$\\
Assume $\tilde{x} \in \mathbb{Re}^n \geq 0$ and solve algebraically for $F(\tilde{x}(t)) = 0$ to give fixed-point solutions, $\tilde{x}^{g,*}$\\
Compute the Jacobian matrix $\mathbf{J}$ \\
Check stability of each fixed-point, $x^{\gamma,*}$, for $\gamma = 1,\ldots,g$: \\
\begin{algorithmic}[1]
\Require{Linear Stability Analysis}
\item Set $x(0)' = x^{\gamma,*} + q$
\item Solve $F(x(t)')$
\item Calculate $u = x(t) - x^{\gamma,*}$
\setlength{\itemindent}{0.3in}
\item[]  \eIf{$||u|| < \epsilon$}{$x^{\gamma,*}$ is stable}{$x^{\gamma,*}$ is unstable}
\end{algorithmic}
\begin{algorithmic}[1]
\normalsize \Require{Routh-Hurwitz Criterion}
 \item Calculate characteristic polynomial of $\mathbf{J}$: $p_\mathbf{J}(\tilde{\lambda}) = \det(\tilde\lambda \mathbf{I} - \mathbf{J})$
\item Calculate the eigenvalues $\tilde{\lambda}$ by finding the roots of $p_\mathbf{J}(\tilde{\lambda})$
\setlength{\itemindent}{0.3in}
\item[]  \uIf{$\forall.\lambda \in \mathbb{Re} (\lambda < 0)$}{$x^{\gamma,*}$ is a stable node}
		\uElseIf{$\forall.\lambda \in \mathbb{Re} (\lambda > 0)$}{$x^{\gamma,*}$ is an unstable node}
		\uElseIf{$\{\exists.\lambda \in \mathbb{Re}(\lambda > 0) | \exists.\lambda \in \mathbb{Re}(\lambda < 0)\}$}{$x^{\gamma,*}$ is an unstable saddle point}
		\uElseIf{$\forall.\lambda \in \Re (\lambda < 0)$}{$x^{\gamma,*}$ is a stable spiral}
		\uElseIf{$\forall.\lambda \in \Re (\lambda > 0)$}{$x^{\gamma,*}$ is an unstable spiral}
		\ElseIf{$\forall.\lambda \in \Re (\lambda = 0)$}{$x^{\gamma,*}$ is a circle}
\end{algorithmic}
\caption{Single-point numerical analysis}
\label{algoa}
\end{algorithm}

\subsubsection*{Multiple-point analysis}
This algorithm is an extension of the single-point routine in which the stability analysis is performed for a range of user-specified values $(a_i,b_i)$, with step-size $s_i$, for any given parameter pair, $\theta_i$ for $i=1,2$. The resulting steady-state regions and their stability are visualised as a two-dimensional bifurcation phase plane for ($\theta_1,\theta_2$). This is a powerful method for understanding the relationship between model parameters and the existence and stability of the system steady-states. In biological terms, the chosen parameters may represent operational properties of the system (e.g., dilution rate $D$, substrate input concentration $S_{\mathrm{in}}$) or of the organisms themselves (e.g., growth rates, substrate yields), and thus can be used to test the effect of these parameters on the behaviour of the system itself. The method is described in Algorithm~\ref{algob}.

\begin{algorithm}
\SetKw{Step}{step}
\KwIn{$\Theta', \{\theta_i \in \mathbb{Re} | \theta_i \geq 0\}, a_i, b_i, s_i$}
\KwOut{$\tilde{x}(t), \tilde{x}^{g,*} \in \mathbb{Re}^n \times \mathbb{Re}^g$ }
Assume $(\tilde{x},\theta_i) \in \mathbb{Re}^n \geq 0$ and solve algebraically for $F(\tilde{x}(t,\theta_i))=0$ to give fixed-point solutions, $\tilde{x}^{g,*}(\theta_i)$ \\
Define the Jacobian matrix $\mathbf{J}$ symbolically\\
\For {$\theta_1:=a_1$ \KwTo $b_1$ \Step $s_1$}{
\For {$\theta_2:=a_2$ \KwTo $b_2$ \Step $s_2$}{Calculate $\tilde{x}^{g,*}$ and $\mathbf{J}(\theta_i,\tilde{x}^{g,*})$\\ Calculate characteristic polynomial of $\mathbf{J}$: $p_\mathbf{J}(\tilde\lambda) = \det(\tilde\lambda \mathbf{I} - \mathbf{J})$\\\normalsize Calculate the eigenvalues $\tilde{\lambda}$ by finding the roots of $p_\mathbf{J}(\tilde{\lambda})$ and determine stability $\mathcal{T}$ by:\\{\uIf{$\forall.\lambda \in \mathbb{Re}(\lambda<0)$}{$\tilde{x}^{g,*}$ is stable }\Else{$\tilde{x}^{g,*}$ is unstable}}
Determine steady-state, $\mathcal{SS}$, from $\{X_n \in \mathbb{Re} | X_n>0 \}$\\
Assign steady-state stability region as $\mathcal{J}_j=\mathcal{SS}\cap \mathcal{T}$}
}
Plot $\mathcal{J}_j(\theta_i)$
\caption{Multiple-point numerical analysis}
\label{algob}
\end{algorithm}

\subsubsection*{Basin of attraction analysis}
This algorithm (Algorithm~\ref{algoc}) is used to determine the basin of attraction for any pair of variables within $\tilde{x}$, by assessing the influence of their initial conditions on the steady-state. Ecological systems can exhibit multiple stable states within the same parameter space. In other words, depending on the initial conditions, a system may tend towards different equilibrium points representing distinct biological behaviour. For example, the concept of resilience extends the purpose of investigating ecological systems beyond the goal of maintaining equilibrium (stability) to one in which system behaviour is reliant on the nature and character of the basins of attraction and its interaction with external factors~\cite{holling73}. In this way, it is informative to determine the effect that the initial conditions have in driving the system to alternate stable states. This allows for practical interpretation of operating and control strategies to ensure a system remains in the desired basin of attraction~\cite{fassoni14}. 

\begin{algorithm}
\SetKw{Step}{step}
\KwIn{$\Theta, \{\tilde{x}(0) \in \mathbb{Re}^n | \tilde{x}(0) \geq 0\}, a_n, b_n, s_n, \epsilon$}
\KwOut{$\tilde{x}(t), \tilde{x}^{g,*} \in \mathbb{Re}^n \times \mathbb{Re}^g$}
Substitute numerical parameters $\Theta$ into system of symbolic ODEs $F(\tilde{x}(t))$ \\
Assume $\tilde{x} \in \mathbb{Re}^n \geq 0$ and solve algebraically for $F(\tilde{x}(t))=0$ to give fixed-point solutions, $\tilde{x}^{g,*}$ \\
Select two variable initial conditions to sweep, $x_{(1,2)}(0)'$, and fix the initial conditions of the other variables, $\tilde{x}(0)$\textbackslash $x_{(1,2)}(0)$\\
\For {$x_1(0):=a_1$  \KwTo $b_1$  \Step $s_1$}{
\For {$x_2(0):=a_2$  \KwTo $b_2$ \Step $s_2$}{Solve $F(\tilde{x}(t)')$\\  \For{$ii:=1$ \KwTo $g$}{Calculate $\tilde{u} = \tilde{x}(t) - \tilde{x}^{*}$\\ \eIf{$||\tilde{u}|| < \epsilon$}{Fixed-point attractor is $\tilde{x}^{*}$}{Fixed-point attractor $\nexists$}Determine steady-state, $\mathcal{SS}$, from $\{X_n \in \mathbb{Re} | X_n>0 \}$\\}}}
Plot $\mathcal{SS}((a,b)_{(1,2)})$
\caption{Basin of attraction analysis}
\label{algoc}
\end{algorithm}

\subsubsection*{Phase portrait}
The phase portrait algorithm (Algorithm~\ref{algod}) is an alternative approach for visualising the location of state attractors. Unlike the basin of attraction analysis, this routine shows the trajectories of the solutions for the system of ODEs. Critically, it can provide important information on the dynamics of the system in two or three dimensions.\\

\begin{algorithm}
\SetKw{Step}{step}
\KwIn{$\Theta, \{\tilde{x}(0) \in \mathbb{Re}^n | \tilde{x}(0) \geq 0\}, a_n, b_n, s_n$}
\KwOut{$\tilde{x}(t)$}
Solve $F(\tilde{x}(t))$ \textbf{for} $a_n \leq \tilde{x}(0) \leq b_n$ \Step $s_n$\\
Plot $ F(\tilde{x}(t))|_{a_n}^{b_n}$
\caption{Calculation of Phase Portrait}
\label{algod}
\end{algorithm}

\subsection*{Thermodynamics module}

Mathematical modelling of microbial growth is typically tethered to empirical observations; kinetic models that are derived from experiments with parameters that may have no biological meaning (e.g., the half-saturation constant $K_S$ is merely an estimated \textit{shaping factor} that is highly sensitive to both biological and physical characteristics of the process being modelled~\cite{arnaldos15}). Metabolic thermodynamics have been proposed to resolve the weaknesses in purely kinetics based modelling, especially in systems with organisms growing under moderately exergonic conditions where a trade-off between kinetic and thermodynamically favourable conditions become apparent (e.g., anaerobic digestion~\cite{grosskopf16}). Whilst kinetic models are restricted by their underlying mechanics, which govern the rate of change in reaction components, thermodynamics is immutable and describes the stability and direction of the same reaction. In this sense, for energy constrained systems, it defines the feasible \textit{free-energy} conditions under which a species may grow. A number of models propose a framework by which single- and multiple-reactant thermodynamics may be integrated with classical growth kinetics~\cite{rodriguez08,gonzalez13,flamholz13}.

We have implemented the simplest approach to incorporating thermodynamics, which only considers replacement of a kinetic inhibition function with one describing the effect of available energy on microbial growth. The function does not describe the thermodynamic energy required for anabolism, i.e. cellular maintenance or biomass production, and is only necessary when considering cases in which the free energy is limiting. We feel that this is an adequate method for implementing thermodynamics in models that describe an average behaviour of a microbial community, rather than individual based models where spatial diversity may also be considered \cite{gonzalez13,hellweger16}. Furthermore, more complex thermodynamic models, although useful in general microbial ecology modelling, may make the analyses provided here intractable. Extension of the thermodynamics is, therefore, not included here. The general form of the growth function using this inhibition term is given by:

\begin{equation}
f_{growth} = \frac{k_{m,i}S_{i}}{K_{S,i} + S_{i}}I_{th}
\end{equation}

\noindent where $k_{m,i}$ is the maximum specific growth rate of species $i$ and:

\begin{equation}
I_{th} = 1-e^{\frac{\Delta G_{rxn}}{RT}}
\end{equation}

\noindent where the Gibbs free energy change in the reaction is given by the standard equation:

\begin{equation}
\Delta G_{rxn} = \Delta G^0{'}+ RT\ln\left(\frac{\prod{S_\pi}}{\prod{S_\nu}}\right)
\end{equation}

\noindent where $\Delta G^0{'}$ is the Gibbs free energy of all unmixed chemical species in the reaction (adjusted for temperature using Van't Hoff's equation),  $S_\pi$ are the reaction products, $S_\nu$ are the reactants/substrates, $R$ is the universal gas constant, and $T$ is the temperature.

This module is available upon selection of the \textit{thermodynamics} option from the  \begin{myfont}MI-SIM\end{myfont} interface. The growth model functions, $f_n$, default to Monod form in all cases. \\

The thermodynamic calculator comprises three steps:\\

\begin{itemize}
\item \textbf{System specification}: The user should enter the operating temperature (K), reaction $\Delta G^0$ (kJ/mol), and the index of reactants ($r$) and products ($p$) in the reaction. $\Delta G^0$ can be set to zero if the user wishes to automatically calculate the $\Delta G$ value for each component of the reaction from the drop-down menus. For compounds not available in these menus, the cumulative $\Delta G$ values must be entered in the $\Delta G^0$ box provided. Although $r$ and $p$ are set to one in all cases, it is necessary here to specify the actual reaction stoichiometry to properly define its thermodynamics. Hence, there may be more reactants or products involved that are not explicitly modelled by \begin{myfont}MI-SIM\end{myfont}. A simple text display of the reaction is provided to indicate the substrates, $S_n'$ and biomass $X_n$ involved in the conversion.

\item \textbf{Compound specification}: In the central block of the calculator, the user may input stoichiometric information for each reactant and product featured in the reaction. This includes the number of moles of substrate and product, the molar mass of each compound (g/mol) and, if required, the option to calculate the Gibbs free energy for each compound. For the latter, the user must specify the compound using drop-down menus to choose the chemical system where it is found, and its state (e.g., solid, liquid, aqueous, gas). For the product compounds, the user must specify their stoichiometric coefficient, $\gamma_p$. It should be noted that the molar mass is only required if the selected system units are mass of COD, whereby an additional conversion term (molarity to units COD, using the molar mass value) is required to calculate the $\Delta G_{rxn}$. 

\item \textbf{Calculation}: The calculation of the thermodynamic inhibition function is executed by pressing the \textit{Calculate} button. The values for the compound specific $\Delta G_n^0$ at temperature $T$ are taken from the $\Delta G^0$ tables~\cite{amend01}. Polynomial curve fitting of order 2 is used to determine the $\Delta G^0$ values for the user-specified temperature, based on the temperature dependent curves developed from the values in the cited tables. \end{itemize}

The \textit{Add Sn to ODE} function automatically defines equations for the additional substrates and products ($S_{\nu,new}$ and $S_{\pi,new}$) not currently included in the selected motif.
In the case additional reactants ($S_{\nu,new}$) are included, a reconfiguration of the growth function ($f_{new}$) for biomass $X_n$ is required to account for its growth on these additional substrate ($\mu_{new}$). The generic form of the additional equations are as follows:\\
\begin{align}
\frac{dS_{\nu,new}}{dt}&=-DS_{\nu,new}-f_{new}X_nI_{th}\\
\frac{dX_n}{dt}&=-DX_n+Y_{new}f_{new}X_nI_{th}-k_{dec,n}X_n\\
\frac{dS_{\pi,new}}{dt}&=-DS_{\pi,new}+\gamma_{new}(1-Y_n)f_nX_nI_{th}\\
f_{new}&=f_n\mu_{new}
\end{align}

The \textit{Finish} button closes the thermodynamic calculator module and returns the user to the main GUI interface.

\subsection*{Example: A three-tiered food-web}
\begin{figure}[h]
\centering
\includegraphics[width=0.8\textwidth,clip=true]{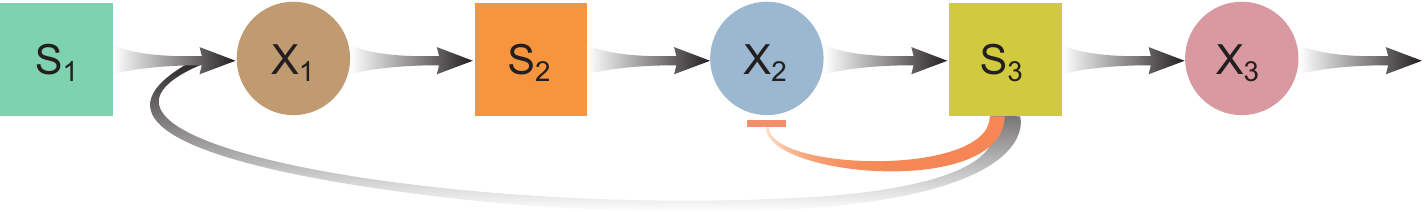}
\caption{A three species food-web with competitive and syntrophic interactions. The black arrows represent substrate/nutrient flow through the system and the orange arrow is the non-competitive inhibition of $X_2$ by the substrate $S_3$.}\label{Fig_1}
\end{figure}

To demonstrate the functionality of \begin{myfont}MI-Sim\end{myfont}, we analysed a three-tiered microbial food-web with competitive and syntrophic interactions, as shown in Fig.~\ref{Fig_1}, and described by Eqs~(\ref{eq:eq1})-(\ref{eq:eq12}), with units expressed in COD. To parameterise the model, we took the values given in a previous analysis performed independently of \begin{myfont}MI-Sim\end{myfont}~\cite{wade16}, with a revised and more realistic value for the substrate affinity constant, $K_{S,1}$, shown in Table~\ref{pval}. The model is described by the six ODEs:\\

\begin{figure}[h]
\centering
\includegraphics[width=0.4\textwidth,clip=true,angle=90]{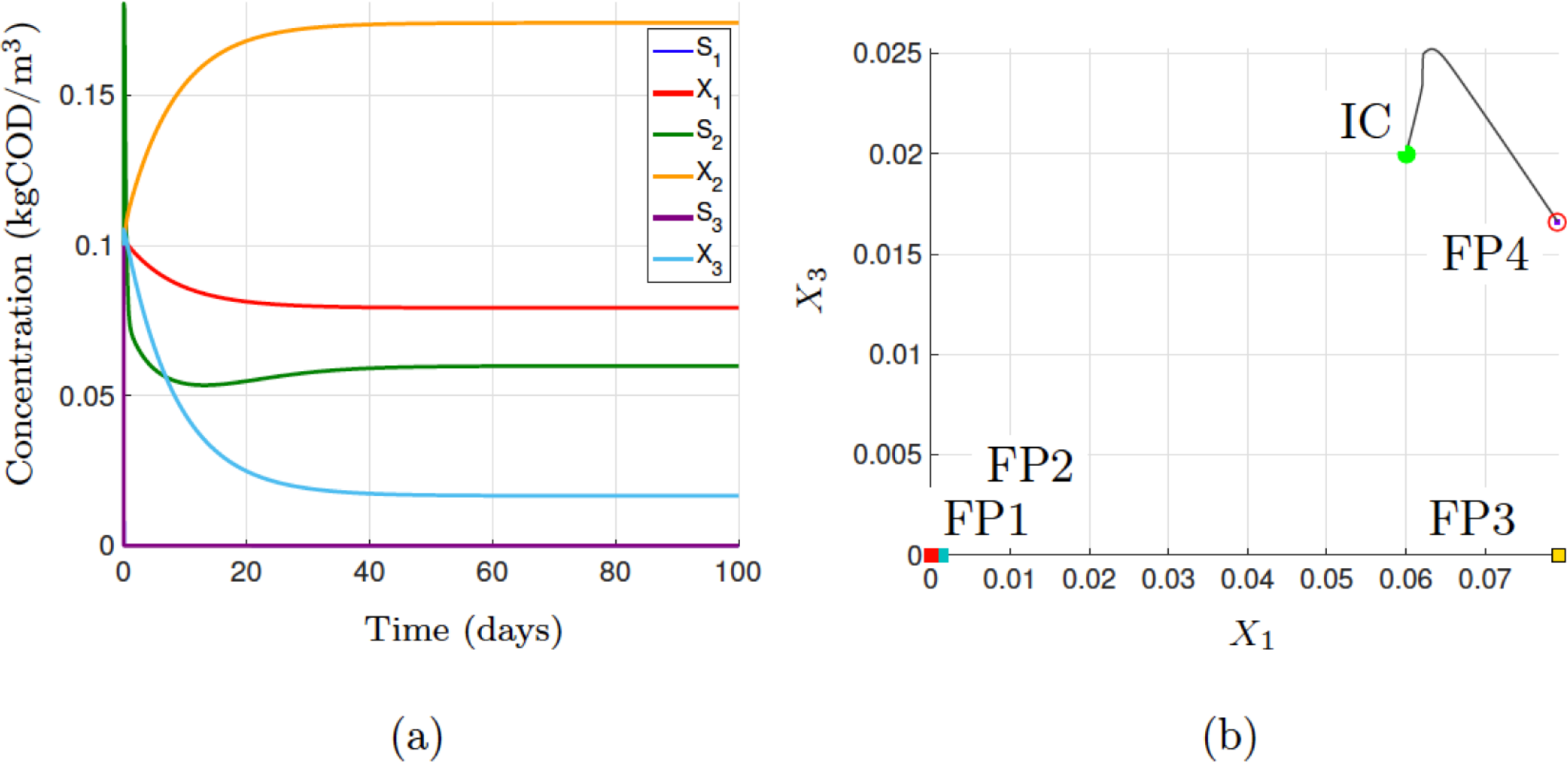}
\caption{ Single-point analysis results showing (a) the three-tiered motif dynamics and (b) a phase portrait of the fixed-point attractors for $X_1$ against $X_3$, with $X_1(0) = 0.06$ and $X_3(0) = 0.02$ kgCOD/m$^3$.}\label{Fig_2}
\end{figure}

\begin{table}[bph!]\caption{Model parameters for the three-tiered food-web example.}
\centering
\begin{tabular}{lll}
\hline 
Parameter & Value & Units\\ \hline
$k_{m,1}$ & 29.12 & kgCOD$_S$/kgCOD$_X$/d\\
$K_{S,1}$ &  5.2$\times10^{-5}$ & kgCOD/m$^3$ \\
$Y_{1}$ & 0.019 & kgCOD$_X$/kgCOD$_S$  \\ 
$k_{\mathrm{dec},1}$ & 0.02 & days$^{-1}$\\
$k_{m,2}$ & 26 & kgCOD$_S$/kgCOD$_X$/d\\
$K_{S,2}$ &  0.302 & kgCOD/m$^3$\\
$Y_{2}$ & 0.04 & kgCOD$_X$/kgCOD$_S$  \\ 
$k_{\mathrm{dec},2}$ & 0.02 & days$^{-1}$ \\
$k_{m,3}$ & 35 & kgCOD$_S$/kgCOD$_X$/d\\
$K_{S,3}$ &  2.5$\times10^{-5}$ & kgCOD/m$^3$\\
$Y_{3}$ & 0.06 & kgCOD$_X$/kgCOD$_S$ \\ 
$k_{\mathrm{dec},3}$ & 0.02 & days$^{-1}$ \\
$K_{i,3}$ & 3.5$\times10^{-6}$ & kgCOD/m$^3$  \\
$\gamma_{0}$ & 1.0769 & -\\
$\gamma_{1}$ &  0.1429 & -\\
$\gamma_{2}$ & 0.0769 & - \\ 
$S_{2,\mathrm{in}}$ & 0 & kgCOD/m$^3$\\
$S_{3,\mathrm{in}}$ & 0 & kgCOD/m$^3$ \\ \hline
\end{tabular}
\label{pval}
\end{table}

\begin{align}
\label{eq:eq1} \frac{dS_1}{dt} &= D(S_{1,\mathrm{in}} - S_1) - f_1X_1\\
\frac{\mathrm{d}X_1}{\mathrm{d}t} &= -DX_1 + Y_1f_1X_1 - k_{\mathrm{dec},1}X_1\\
\frac{\mathrm{d}S_2}{\mathrm{d}t} &= D(S_{2,\mathrm{in}} - S_2) + \gamma_0(1-Y_1)f_1X_1 - f_2X_2I_3\\
\frac{\mathrm{d}X_2}{\mathrm{d}t} &= -DX_2 + Y_2f_2X_2I_3 - k_{\mathrm{dec},2}X_2\\
\frac{\mathrm{d}S_3}{\mathrm{d}t} &= D(S_{3,\mathrm{in}}-S_3) + \gamma_1(1-Y_2)f_2X_2I_2 - f_3X_3 - \gamma_2f_1X_1\\
\frac{\mathrm{d}X_3}{\mathrm{d}t} &= -DX_3 + Y_3f_3X_3 - k_{\mathrm{dec},3}X_3
\end{align}

where:
\begin{align}
f_1 &= \frac{k_{m,1}S_1}{K_{S,1} + S_1}\frac{S_3}{K_{S,3c} + S_3}\\
f_2 &= \frac{k_{m,2}S_2}{K_{S,2} + S_2}\\
f_3 &= \frac{k_{m,3}S_3}{K_{S,3} + S_3}\\
\label{eq:eq12} I_3 &= \frac{1}{1+\frac{S_3}{K_{i,3}}}
\end{align}
We selected values of two operating parameters as dilution rate, $D=0.01 \ \mathrm{d}^{-1}$, such that $D < \max(k_{m,n}Y_{m,n})$, and substrate input concentration, $S_{1,\mathrm{in}} = 5 \ \mathrm{kgCOD/m}^3$. 
Arbitrarily setting the initial conditions for the six variables to $0.1$ and selecting the stiff ODE solver \textit{odes23s} with default tolerances, we ran the single-point analysis for 100 days and observed that steady-state was reached after approximately 70 days, as shown in Fig.~\ref{Fig_2}a. A trajectory plot showing the dynamical relationship between species $X_1$ and  $X_3$ is shown in Fig.~\ref{Fig_2}b. These plots can be compared to the results of the fixed-point analysis, which are displayed in the GUI window, and presented here in Table~\ref{fpa}. The results show that four fixed points (FP), $\tilde{x}^{g,*} \geq 0$, are possible for the given parameters: i) complete washout of the three species (FP1), ii) washout of species $X_3$ (FP2/FP3), and iii) maintenance of all three species (FP4). Given the initial conditions (ICs), the steady-state reached is case iii), which is verified by examining both of the figures.

\begin{table}[bph!]\caption{Fixed-point and stability analysis results for the example considered.}
\centering
\small
\begin{tabular}{cccccccc}
\hline 
Fixed-point & $S_1$ & $X_1$ & $S_2$ & $X_2$ & $S_3$ & $X_3$ & Stability \\ \hline
$\mathrm{FP1}$ & $5$ & $0$ & $0$ & $0$ & $0$ & $0$ & Stable spiral \\
$\mathrm{FP2}$  & $4.91$ & $1.37\times10^{-3}$ & $0.04$ & $1.62\times10^{-3}$ & $2.77\times10^{-7}$ & $0$ & Unstable \\
$\mathrm{FP3}$ & $1.51\times10^{-5}$ & $0.08$ & $2.48$ & $0.09$ & $2.35\times10^{-5}$ & $0$ & Unstable \\
$\mathrm{FP4}$  & $2.93\times10^{-5}$ & $0.08$ & $0.06$ & $0.17$ & $1.52\times10^{-6}$ & $0.02$ &  Stable node  \\ \hline
\end{tabular}
\label{fpa}
\end{table}

\begin{figure}[h]
\centering
\includegraphics[width=0.4\textwidth,clip=true,angle=90]{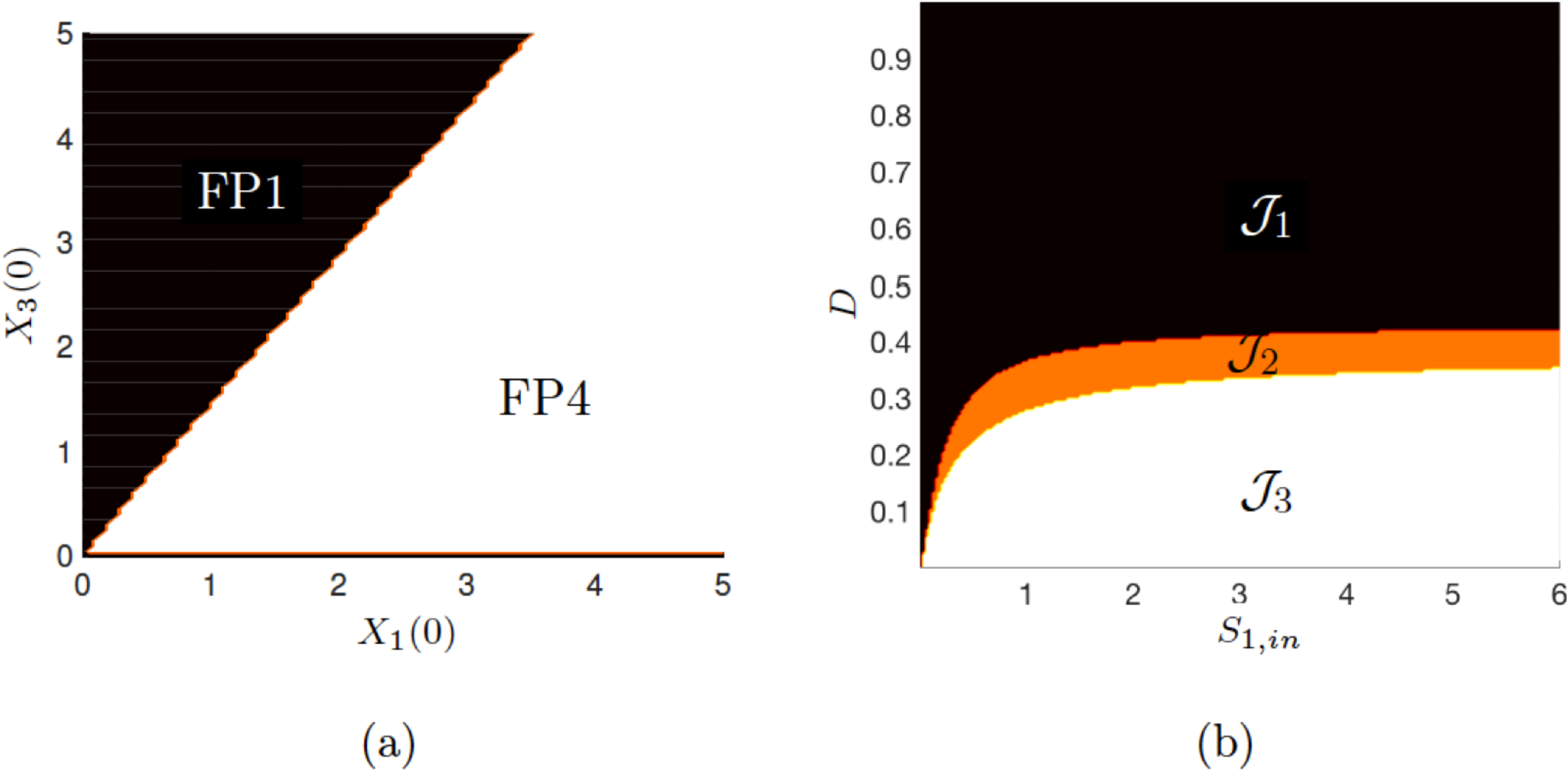}
\caption{Output from multiple-point algorithms showing (a) basin of attraction analysis varying the initial conditions for $X_1$ and $X_3$, and (b) multiple-point steady-state existence and stability of the three-tiered motif.}\label{Fig_3}
\end{figure}

Additionally, stability analysis indicates that the system is bistable (FP1 and FP4) for the selected operating parameters, and the \textit{basin of attraction} routine is subsequently used to determine which attractor (fixed-point) the system will converge to, given a range of ICs. This is shown in Fig.~\ref{Fig_3}a, in which the influence of changing the ICs for $X_1$ and $X_3$ is observed. The result indicates that there is a critical ratio for the two initial biomass concentrations ($r_{1,3} = X_1(0)/X_3(0)$) defining a boundary between FP1 and FP4. If $r_{1,3} \geq 0.707$ and $\{X(0)\in \mathbb{Re}^{1,3} | (X_1(0),X_3(0)) > 0\}$, then $\mathcal{SS}3$ (steady-state 3 $\equiv$ FP4) is assured. It should be noted that steady-states may comprise of one or two fixed-points in this example. \\

We also performed the analysis for multiple operating points ($D, S_{1,\mathrm{in}}$). This enables the user to visualise the transitions (bifurcations) between steady-state regions and their stability. It can be used to show the theoretical operating space necessary to attain a desired objective, such as community stability. In this example, three regions are identified corresponding to the fixed-points FP1 - FP4. These regions also indicate the stability of the system at each point in the two-dimensional operating space and, together, we denote these regions as $\mathcal{J}_1, \mathcal{J}_2$ and $\mathcal{J}_3$, as shown in Fig.~\ref{Fig_3}b and Table~\ref{jreg}. It is observed that $\mathcal{SS}1 \equiv$FP1 is always stable, whilst $\mathcal{SS}2$ is associated with two fixed points (FP2, FP3) where it exists, one of which is always unstable. 

\begin{table}[bph!]\caption{Multiple-point stability analysis.}
\centering
\begin{tabular}{cccc}
\hline 
Region & $\mathcal{SS}1$&$\mathcal{SS}2$&$\mathcal{SS}3$\\ \hline
$\mathcal{J}_1$ & Stable & &\\
$\mathcal{J}_2$ & Stable & Unstable/Stable & \\
$\mathcal{J}_3$ & Stable & Unstable/Unstable & Stable  \\ \hline
\end{tabular}
\label{jreg}
\end{table}

\subsection*{Limitations and future work}
\subsubsection*{Computational Performance}
\begin{myfont}MI-Sim\end{myfont} employs a number of algorithms that may be computationally expensive. This can be an issue in \begin{myfont}MATLAB\end{myfont}, especially when nested loops are used. We have attempted to overcome performance concerns by using vectorisation instead of loops wherever possible and by making use of \begin{myfont}MATLAB\end{myfont} specific commands such as \textit{matlabFunction} for handling symbolic manipulation. In the case where loops are unavoidable, increasing the resolution of the output by specifying larger numbers of calculation points will significantly impact performance. This is most relevant for higher-dimensional models, such as the three-tiered motif we include in the software. This is a challenging system to solve using \begin{myfont}MATLAB\end{myfont}, so we employ \begin{myfont}MuPad\end{myfont}, which has a powerful symbolic engine that can solve complex systems of ODEs to generate all possible solutions. In \begin{myfont}MATLAB\end{myfont}, the solutions may be incomplete and overriding the constraints for such systems results in failure. The \begin{myfont}MuPad\end{myfont} option is computationally costly, but once solutions are found, subsequent analysis can be parallelised to significantly improve performance.

\subsubsection*{Dimensionality}
The current version of \begin{myfont}MI-Sim\end{myfont} allows for the analysis of seven motifs comprised of two or three species, described by up to six ODEs. Theoretically, \begin{myfont}MATLAB\end{myfont} can solve higher-dimensional systems of ODEs numerically, but this is dependent on the memory available to the CPU. For stiff problems, ODE solvers allocate memory for solution vectors of length $n$ and a Jacobian matrix of size $n\times n$. It is recommended to use the \textit{Jacobian} solver option for these cases. The restriction to systems of six ODEs or less is necessary to ensure that the software will run on most standard CPUs, however experienced modellers may wish to test the algorithms with larger models. Future development of the software will aim to allow users to enter their equations manually via the interface and automatically configure the parameter, variable and solver options accordingly. Further, greater access to solver options and dimension reduction techniques may overcome some of the memory limitations of \begin{myfont}MI-Sim\end{myfont}.

\section*{Software implementation}
\begin{myfont}MI-Sim\end{myfont} has been developed in \begin{myfont}MATLAB\end{myfont} (R2015a, version 8.5.0) and tested on both R2016a and R2016b. In order to facilitate the use of the software and to motivate its further development, the \begin{myfont}MI-Sim\end{myfont} package and documentation have been released at \url{http://mi-sim.github.io}. The software is also available for direct download on the Mathworks File Exchange at \url{http://uk.mathworks.com/matlabcentral/fileexchange/55492-mi-sim}.

\section*{Conclusion}
We have developed the \begin{myfont}MATLAB\end{myfont} based software tool \begin{myfont}MI-Sim\end{myfont} for numerical analysis of ecological interactions. The tool is built on a system of ODEs encompassing three to six dimensions. \begin{myfont}MI-Sim\end{myfont} allows for the rigorous analysis of a range of standard two species interaction motifs as well as an extended motif of three species. As well as profiling of the basic dynamics of the motif for a given set of parameters, \begin{myfont}MI-Sim\end{myfont} also provides the capability to analyse the system in a two-dimensional parameter space to enable investigation of system properties such as existence and stability of steady-states and bifurcation trajectories. In cases where the motif contains multi-stable regions, the basin of attraction tool is useful in understanding the role of the initial conditions on the ultimate equilibrium condition. Future releases of \begin{myfont}MI-Sim\end{myfont} will include the possibility of including user-defined motifs for analysis, as well as a sensitivity analysis tool to explore the effect of parameter uncertainty on the behaviour and properties of the system.

\section*{Data accessibility}
The data provided here can be generated directly from open-access the software provided with this article. 

\section*{Acknowledgments}
This work was funded by the Biotechnology and Biological Sciences Research Council UK (BB/K003240/2 Engineering synthetic microbial communities for biomethane production) and by the Institute of Sustainability, Newcastle University. The authors would like to thank George Stagg at Newcastle University for his help in setting-up the github repository.

\end{document}